\documentclass[a4paper,11pt]{article}
\usepackage{pos}
\usepackage{float}
\title{Charged Higgs Boson Mass Bounds in 2HDM-II: Impact of Vector-Like Quarks}

\author[a]{Rachid Benbrik}
\author*[a]{Mohammed Boukidi}
\author[b,c]{Stefano Moretti}

\affiliation[a]{Polydisciplinary Faculty, Laboratory of Fundamental and Applied Physics, Cadi Ayyad University,\\ Sidi Bouzid, B.P. 4162, Safi, Morocco}
\affiliation[b]{Department of Physics \& Astronomy, Uppsala University,\\ Box 516, SE-751 20 Uppsala, Sweden}
\affiliation[c]{School of Physics \& Astronomy, University of Southampton,\\ Southampton, SO17 1BJ, United Kingdom}
\emailAdd{r.benbrik@uca.ac.ma}
\emailAdd{mohammed.boukidi@ced.uca.ma}
\emailAdd{stefano.moretti@physics.uu.se}
\emailAdd{s.moretti@soton.ac.uk}

\abstract{We explore the phenomenology of charged Higgs bosons ($H^\pm$) and Vector-Like Quarks (VLQs), specifically the top-like $T$, within the Two Higgs Doublet Model Type-II (2HDM-II). We consider both a singlet VLQ $(T)$ and a doublet $(TB)$ scenario, demonstrating that the presence of VLQs  influences the scalar sector, particularly by alleviating the stringent mass constraints on $H^\pm$ imposed by $B$-physics observables such as $B\to X_s\gamma$. This relaxation arises from modifications in the couplings between $H^\pm$ and Standard Model (SM) quarks, with the magnitude of the effect differing between the singlet and doublet cases. We further analyse the constraints from the oblique parameters $S$ and $T$ on VLQ mixing angles.}

\FullConference{42nd International Conference on High Energy Physics (ICHEP2024)\\
18-24 July 2024\\
Prague, Czech Republic\\}

\begin{document}
\maketitle

\section{Introduction}
\noindent The discovery of a charged spin-0 boson like $H^\pm$ would provide compelling evidence of physics beyond the Standard Model (BSM), as no such particle exists within the SM, which only includes spin-1 charged bosons and a chargeless spin-0 boson. Recent searches for $H^\pm$ at the LHC have yet to find a signal, leading to strict constraints on BSM scenarios, particularly the Two Higgs Doublet Model Type-II (2HDM-II). In the 2HDM-II, indirect limits from $B$-physics observables, such as $B \to X_s \gamma$, set a lower bound on the charged Higgs mass of around 580 GeV, largely independent of other parameters.

\noindent However, this constraint can be lifted if additional particles are introduced into the 2HDM-II, similar to the Minimal Supersymmetric Standard Model (MSSM), where new particles cancel $H^\pm$ contributions. This paper explores whether introducing Vector-Like Quarks (VLQs), heavy spin-1/2 states with identical electroweak quantum numbers for their left- and right-handed components, can achieve the same effect. VLQs are predicted in many BSM frameworks, such as models with gauged flavour groups, non-minimal supersymmetric scenarios, and composite Higgs models. If their masses are accessible at the LHC, they can be searched for in various final states, with some already being explored by the ATLAS and CMS collaborations.

\noindent Here, we extend the 2HDM-II with singlet $(T)$ and doublet $(TB)$ VLQs in a simplified approach \cite{Arhrib:2024dou, Arhrib:2024tzm}. We demonstrate that VLQ loops in $b \to s \gamma$ transitions can cancel large contributions from $H^\pm$, reducing the lower mass limit of $m_{H^\pm}$ to approximately 200 GeV in the doublet scenario and 500 GeV in the singlet scenario \cite{Benbrik:2022kpo}. However, this reduction requires large VLQ mixing angles, which are constrained by the oblique parameters $S$ and $T$.

\section{Model Descriptions}
\noindent We briefly outline the 2HDM-II+VLQ  framework relevant to this work (for more details, see Refs. \cite{Benbrik:2022kpo, Arhrib:2024dou, Arhrib:2024tzm}). The scalar potential of the $\mathcal{CP}$-conserving 2HDM, with a softly broken $\mathbb{Z}_2$ symmetry via dimension-2 terms, is expressed as:

\begin{equation}
\mathcal{V} = m^2_{11}\Phi_1^\dagger\Phi_1 + m^2_{22}\Phi_2^\dagger\Phi_2 - \left(m^2_{12}\Phi_1^\dagger\Phi_2 + {\rm h.c.}\right) + \text{quartic terms}.
\end{equation}

\noindent The 2HDM can be described in terms of seven independent parameters: $m_h$, $m_H$, $m_A$, $m_{H^\pm}$, $\tan\beta$, $\sin(\beta - \alpha)$, and $m^2_{12}$, which control flavour constraints. Among the four Yukawa types of the 2HDM, we focus on Type-II, where $\Phi_1$ couples to down-type quarks and charged leptons, while $\Phi_2$ couples to up-type quarks.

\noindent For VLQs, we consider singlet $(T)$ and doublet $(TB)$ representations. The $T$ quark carries a charge of $+2/3$, while the $B$ quark has a charge of $-1/3$. The mixing between VLQs and third-generation SM quarks is constrained by low-energy observables, including electroweak precision tests such as the $S$ and $T$ parameters. These constraints primarily impact the properties of top and bottom quarks through radiative corrections.

\noindent The Yukawa Lagrangian for VLQs interacting with third-generation quarks is given by:

\begin{equation}
-\mathcal{L} \, \supset \, y^u \bar{Q}^0_L \tilde{H}_2 u^0_R +  y^d \bar{Q}^0_L H_1 d^0_R + M^0_u \bar{u}^0_L u^0_R  + M^0_d \bar{d}^0_L d^0_R + \rm {h.c.}.
\end{equation}

\noindent The mixing angles between the top quark ($t$) and VLQ ($T$) depend on the VLQ representation. In the singlet case, only $T$ mixes with $t$, while in the doublet scenario, both $T$ and $B$ mix with the third-generation quarks.

\section{Theoretical and Experimental Bounds}
\noindent To validate the framework, we apply a variety of theoretical and experimental constraints. Theoretical requirements include unitarity, perturbativity, and vacuum stability. For electroweak precision observables (EWPOs), the $S$ and $T$ parameters \cite{Benbrik:2022kpo,Abouabid:2023mbu} are computed using dimensional regularization and are compared to global fits at the 95\% confidence level (CL), utilizing \texttt{2HDMC-1.8.0}~\cite{2hdmc}.

\noindent On the experimental side, we impose constraints from SM-like Higgs properties using \texttt{HiggsSignal-3}, limits from collider searches with \texttt{HiggsBounds-6} \cite{Bahl:2022igd}, and constraints from flavour physics observables such as  $B \to X_s \gamma$, $B^+ \to \tau^+ \nu_\tau$, and $B_s \to \mu^+ \mu^-$, using \texttt{SuperIso\_v4.1} \cite{superIso}. 

\section{Numerical Results}\label{sec:results}
\noindent This section provides the numerical results for flavour observables and EWPOs exclusion limits within the 2HDM-II+VLQ framework, focusing on the $(T)$ and $(TB)$ representations of VLQs.

\noindent The most stringent $B$-physics observable we examine is the $b \to s\gamma$ transition. The interactions of the charged Higgs boson with third-generation fermions, which play a critical role in these transitions, are governed by the following Lagrangian:
\begin{equation} -\mathcal{L}{H^+} = \frac{\sqrt{2}}{v}\overline{t}\left(\kappa{t}m_{t}P_L - \kappa_{b}m_{b}P_R\right)bH^+ + {\rm h.c.}, 
\end{equation}
\noindent where $P_{L/R} = (1 \pm \gamma^5)/2$ are the chiral projection operators. The couplings $\kappa_t$ and $\kappa_b$ differ across the 2HDM-II, 2HDM-II+$(T)$, and 2HDM-II+$(TB)$ models, as shown in Table~\ref{coulping}. These couplings are crucial in determining modifications to flavour observables, such as $b \to s\gamma$ and $B_{s/d} \to \mu^+\mu^-$, in different VLQ scenarios.
\begin{table}[H]
	\centering
	{\renewcommand{\arraystretch}{0.6}
		{\setlength{\tabcolsep}{0.05cm}
			\begin{tabular}{c||c||c}
				\hline\hline
				Models & $\kappa_{t}$ & $\kappa_{b}$ \\
				\hline\hline
				2HDM-II & $\cot\beta$ & $-\tan\beta$ \\
				
				2HDM-II+$(T)$ & $c_L \cot\beta$ & $-c_L \tan\beta$ \\
				
				2HDM-II+$(TB)$ & $\cot\beta\left[c_L^d c_L^u + \frac{s_L^d}{s_L^u}(s_L^u{}^2 - s_R^u{}^2) e^{i(\phi_u - \phi_d)}\right]$ & $-\tan\beta\left[c_L^u c_L^d + \frac{s_L^u}{s_L^d}(s_L^d{}^2 - s_R^d{}^2) e^{i(\phi_u - \phi_d)} \right]$ \\
				\hline\hline
	\end{tabular}}}
	\caption{Yukawa couplings of the charged Higgs boson $H^\pm$ to third-generation quarks in 2HDM-II and 2HDM-II+VLQ scenarios.}\label{coulping}
\end{table}

\noindent The Wilson coefficients $C_7$ and $C_8$, governing $b \to s\gamma$ transitions, are proportional to $\kappa_i \kappa_j^*$, with leading contributions expressed as:

\begin{equation}
C_i^{t,\mathrm{model}} = \kappa_b \kappa_{t}^* C_{i,\kappa_{b} \kappa_{t}^*}^{t,\mathrm{model}} + \kappa_t \kappa_{t}^* C_{i,\kappa_{t} \kappa_{t}^*}^{t,\mathrm{model}}.
\end{equation}

\noindent In addition to flavour physics, constraints from EWPOs, particularly the $S$ and $T$ parameters, place limits on VLQ mixing angles. These constraints are essential for determining the allowed parameter space for both $(T)$ and $(TB)$ representations. The scanned parameter ranges for the 2HDM-II and VLQ scenarios are shown in Table~\ref{parameters}.
\begin{table}[H]
	\centering
	{\renewcommand{\arraystretch}{0.2} 
		{\setlength{\tabcolsep}{2.75cm}
			\begin{tabular}{c  c}
				\hline\hline
				Parameters  & Scanned ranges \\
				\hline\hline
				
				\multicolumn{2}{c}{2HDM} \\\hline\hline						
				$m_H$  & [130, 800] \\
				$m_A$  & [80, 800] \\
				$m_{H^\pm}$  & [80, 800] \\
				$\tan\beta$ & [0.5, 20] \\
				$\sin(\beta-\alpha)$ & 1 \\
				\hline\hline
				\multicolumn{2}{c}{ 2HDM-II+$(T)$ } \\\hline\hline
				$s_L$  & [$-$0.5, 0.5] \\
				$m_{T}$   & [750, 2600] 	
				\\\hline\hline
				\multicolumn{2}{c}{2HDM-II+$(TB)$ } \\\hline\hline
				$s_{R}^{u,d}$  & [$-$0.5, 0.5] \\
				$m_{T}$   & [750, 2600] \\	
				\hline\hline
	\end{tabular}}}
	\caption{Parameter ranges for 2HDM and VLQ scenarios. Masses are in GeV. The lightest $\mathcal{CP}$-even Higgs boson $h$ is taken to be around 125 GeV. }
	\label{parameters}
	\end{table}

\subsection{2HDM-II+$(T)$} 
\begin{figure}[H] \centering \includegraphics[width=0.9\textwidth,height=0.4\textwidth]{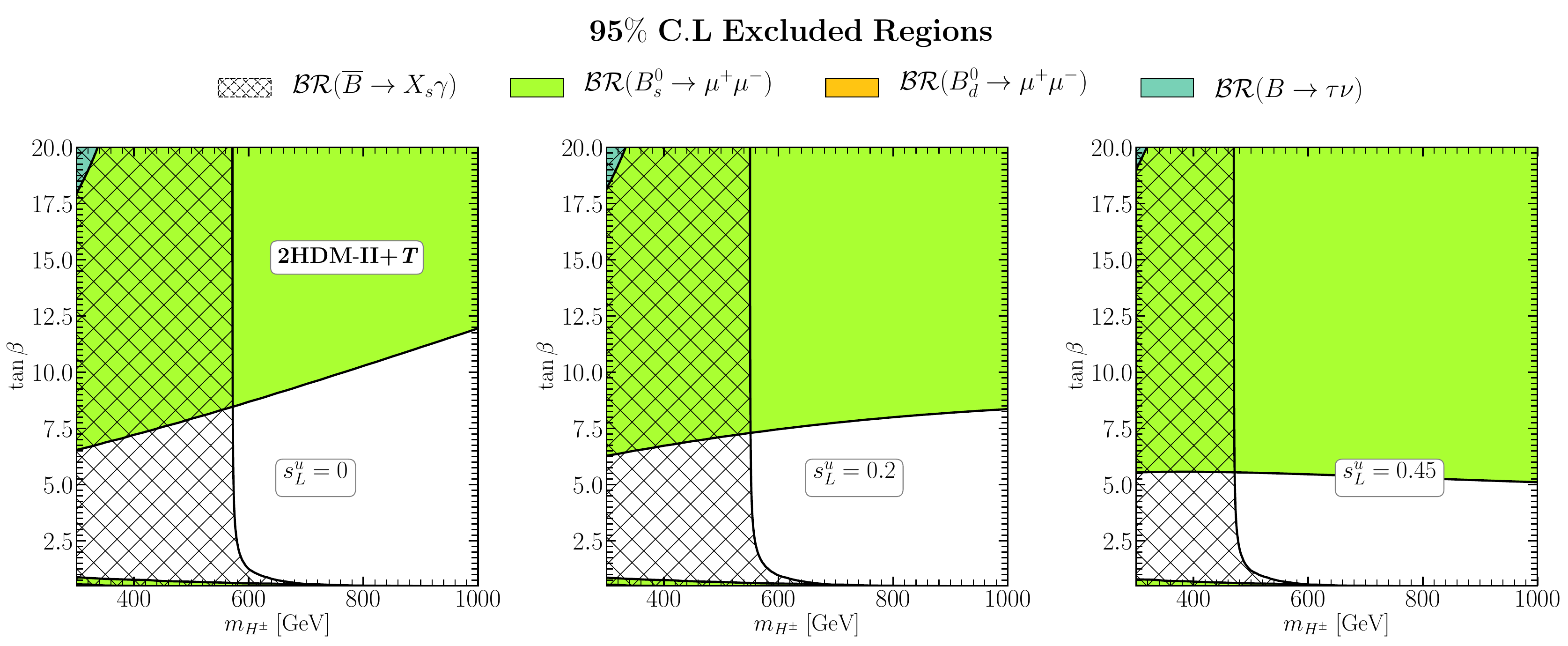} \caption{Excluded regions of the $(m_{H^\pm}, \tan\beta)$ plane by flavor constraints at 95\% CL for the 2HDM-II+$(T)$ singlet with different $s_L^u$ values.} \label{fig1} \end{figure}

\noindent Fig.~\ref{fig1} shows the excluded regions in the $(m_{H^\pm}, \tan\beta)$ plane at 95\% CL based on flavour constraints, specifically from $\bar{B} \to X_s\gamma$, $B_d^0 \to \mu^+ \mu^-$, $B_s^0 \to \mu^+ \mu^-$, and $B_u \to \tau\nu$. As the mixing angle $s_L$ increases, the constraints from $B \to X_s \gamma$ become less stringent, allowing the charged Higgs mass, $m_{H^\pm}$, to drop to approximately 500 GeV for $s_L = 0.45$. However, the limits from $B_s \to \mu^+ \mu^-$ tighten with increasing $s_L$, excluding values of $\tan\beta$ below 7 for $s_L = 0.25$ and below 5 for $s_L = 0.45$.

\begin{figure}[H] \centering \includegraphics[height=5.5cm,width=6.25cm]{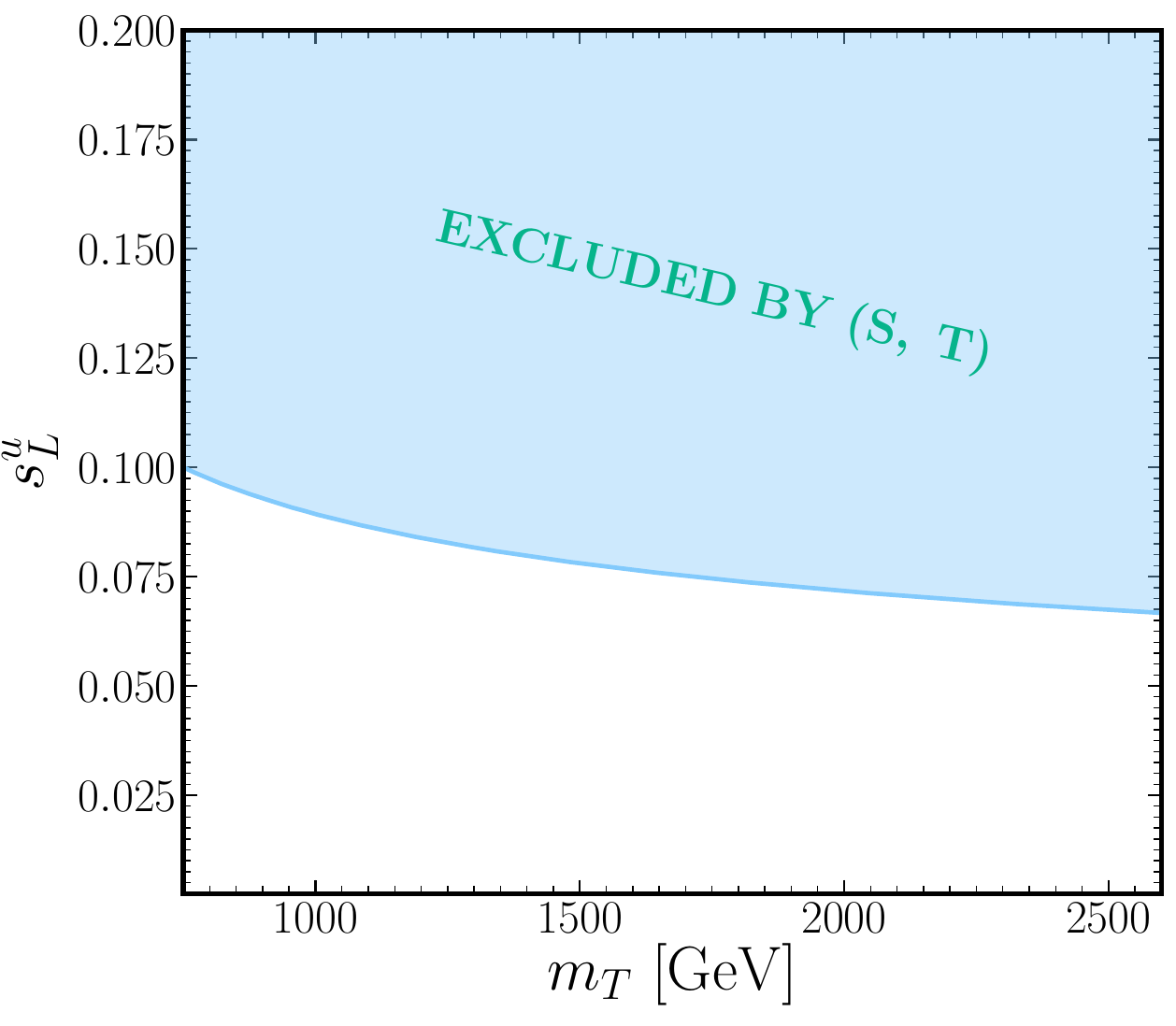} \caption{Exclusion regions in the $(m_T, s^d_R)$ plane based on $S$ and $T$ parameters at 95\% CL, with fixed parameters: $m_{H}=532.80$ GeV, $m_A=524.26$ GeV and $m_{H^\pm}=588.02$ GeV, with $\tan\beta=5.19$.} \label{fig3} \end{figure}

\noindent Fig.~\ref{fig3} illustrates the exclusion regions based on the $S$ and $T$ parameters in the $(m_T, s^u_L)$ plane at 95\% CL. As the VLQ mass $m_T$ increases, the constraints become more stringent, limiting $s^u_L$ to values below 0.1.

\subsection{2HDM-II+$(TB)$} 
\begin{figure}[H] \centering \includegraphics[width=0.9\textwidth,height=0.4\textwidth]{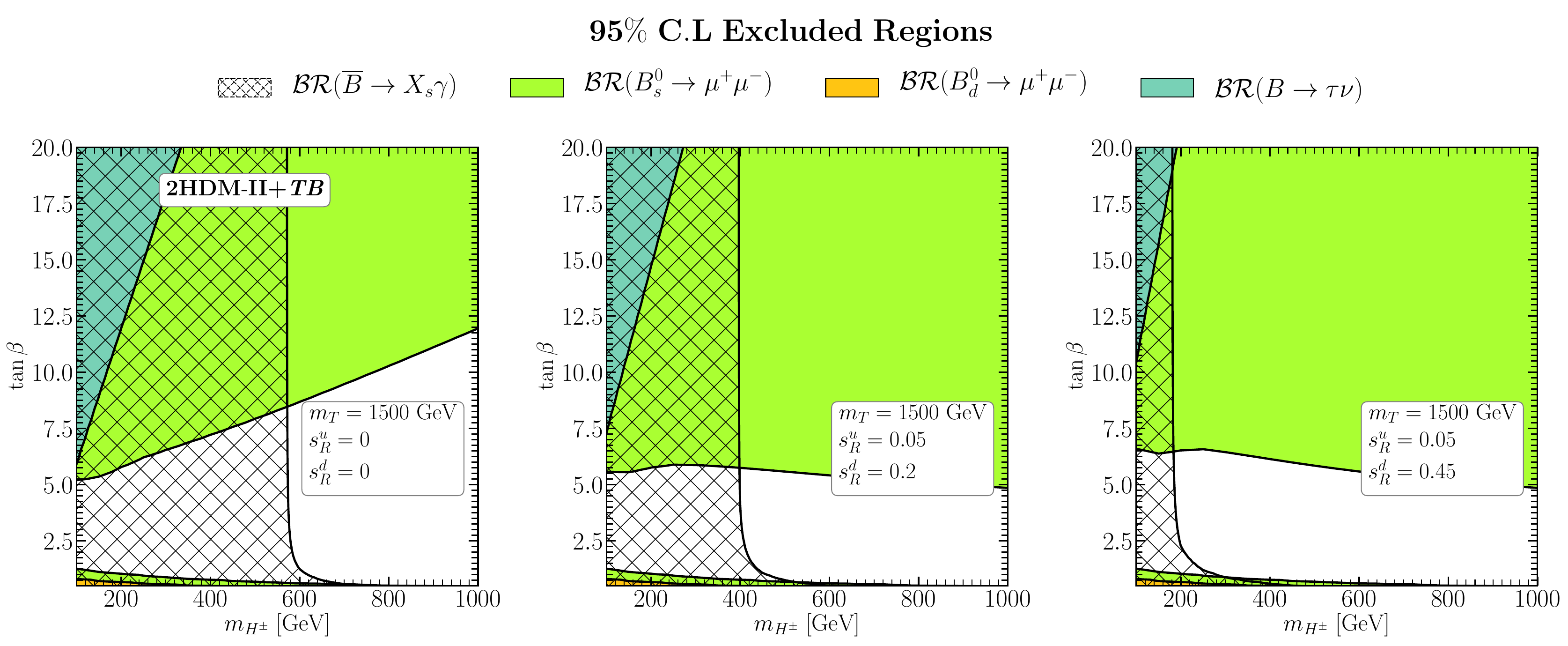} \caption{Excluded regions of the $(m_{H^\pm}, \tan\beta)$ plane by flavor constraints at 95\% CL for the 2HDM-II+$(TB)$ doublet with different $s_R^d$ values.} \label{s^d_R} \end{figure}

\noindent In the 2HDM-II + $(TB)$ scenario, the mixing angles for both up- and down-type quarks significantly influence the model's behaviour. The relation between the mass eigenstates and mixing angles is defined by:

\begin{equation} m_B^2 = \frac{m_T^2 \cos^2\theta_R^{t} + m_t^2 \sin^2\theta_R^{t} - m_b^2 \sin^2\theta_R^{b}}{\cos^2\theta_R^{b}}. 
\end{equation}
\noindent Fig.~\ref{s^d_R} presents the excluded regions in the $(m_{H^\pm}, \tan\beta)$ plane at 95\% CL. As the mixing angle $s_R^d$ increases, the exclusion regions for $B \to X_s \gamma$ shift toward lower values of $m_{H^\pm}$, reaching around 400 GeV for $s_R^d = 0.2$ and approximately 200 GeV for $s_R^d = 0.45$. However, the constraints from $B_s \to \mu^+ \mu^-$ remain stringent, excluding all values of $\tan\beta$ above 5.

\begin{figure}[H] 
	\centering \includegraphics[height=6.75cm,width=11cm]{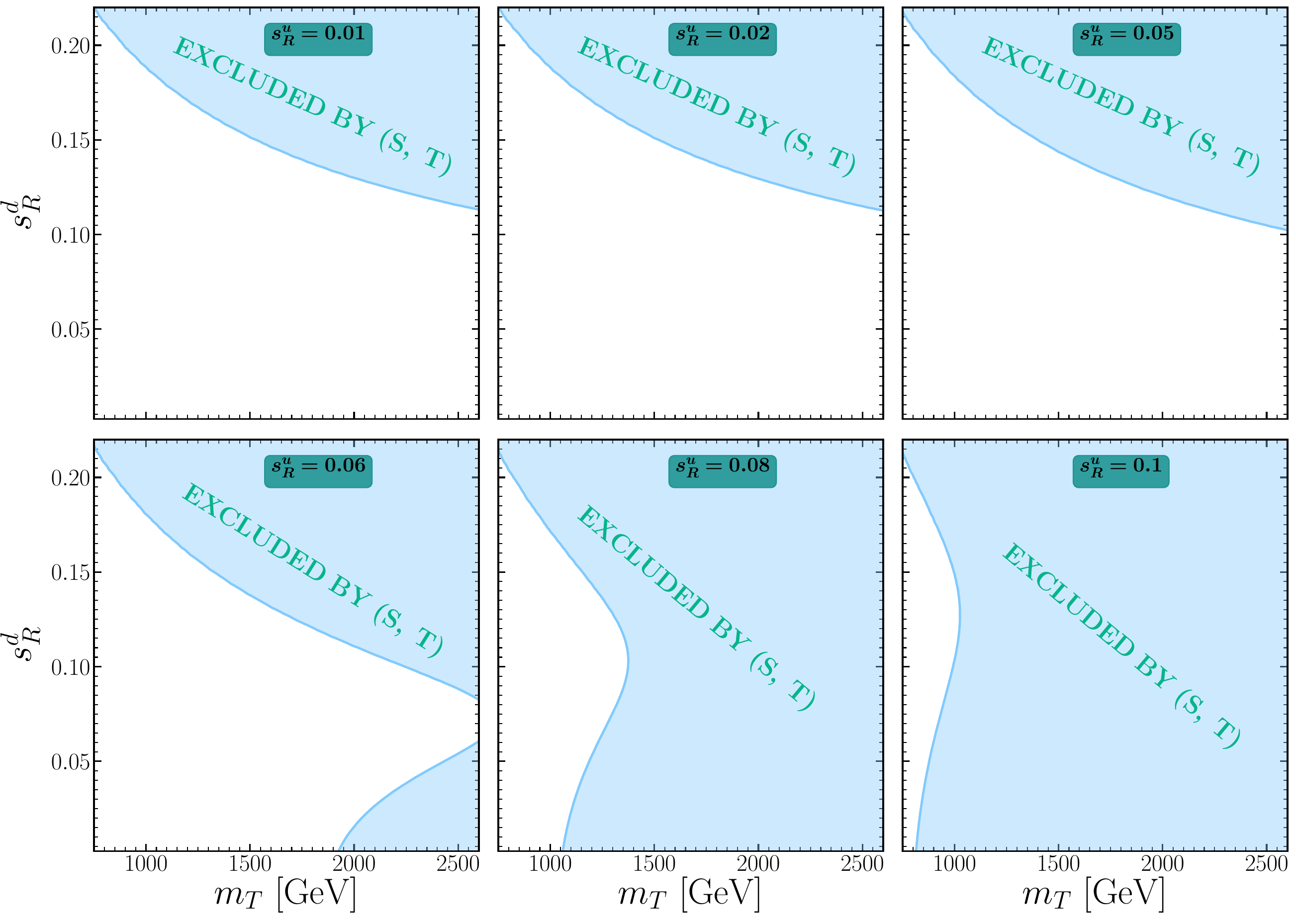}
	 \caption{Exclusion regions by $(S,~T)$ in the $(m_T,~s^d_R)$ plane for various values of $s^u_R$, with fixed parameters: $m_{H}=532.80$, $m_A=524.26$ and $m_{H^\pm}=588.02$, with $\tan\beta=5.19$} \label{fig9} 
\end{figure}

\noindent Fig.~\ref{fig9} highlights the exclusion regions derived from the $S$ and $T$ parameters at the 2$\sigma$ CL. As $s^u_R$ increases, the exclusion region broadens, covering a wider range of $m_T$. A similar expansion is observed with increasing $s^d_R$ and VLQ mass.

\section{Conclusion}
\noindent We examined the effect of VLQs in the 2HDM-II framework, focusing on the $(T)$ (singlet) and $(TB)$ (doublet) representations. Our results demonstrate that VLQs can lower the stringent $m_{H^\pm}$ limits from $B \to X_s \gamma$, reducing the mass to around 500 GeV in the singlet case and as low as 200 GeV in the doublet scenario with large mixing angles. However, oblique parameters $S$ and $T$ impose limits on the mixing angles, constraining this reduction. As a result, the mass limit drops to about 567 GeV for the $(T)$ scenario and 360 GeV for the $(TB)$ scenario, compared to the typical 580 GeV in the 2HDM-II.

\noindent{\bf Acknowledgments}

\noindent
SM is supported in part through the NExT Institute and the STFC Consolidated Grant No. ST/L000296/1.

\end{document}